\documentclass{PoS}

\usepackage{bm}

\title{Differentiable probabilistic programming for strong gravitational lensing}

\ShortTitle{Differentiable probabilistic programming for strong gravitational lensing}

\author{\speaker{Marco Chianese}\thanks{I thank Christoph Weniger for useful comments. I acknowledge the JSPS KAKENHI Grant Number JP18H04340 that partially covered the costs to attend the ICRC 2019 conference.}\\
        Gravitation Astroparticle Physics Amsterdam (GRAPPA), Institute for Theoretical Physics Amsterdam and Delta Institute for Theoretical Physics, University of Amsterdam, Science Park 904, 1098 XH Amsterdam, The Netherlands\\
        E-mail: \email{m.chianese@uva.nl}}

\abstract{The difficult task of observing Dark Matter subhaloes is of paramount importance since it would constrain Dark Matter particle properties (cold or warm relic) and confirm once again the longstanding $\Lambda$CDM model. In the near future the new generation of ground and space surveys will observe thousands of strong gravitational lensing systems providing a unique probe of Dark Matter substructures. Here, we describe a new strong lensing analysis pipeline that combines deep Convolutional Neural Networks with physical models and exploits traditional sampling techniques such as Hamiltonian Monte Carlo. Using simulated strong gravitational lensing systems, we discuss first results and characterize the accuracy of the reconstruction of the main lensing parameters.}

\FullConference{36th International Cosmic Ray Conference -ICRC2019-\\
		July 24th - August 1st, 2019\\
		Madison, WI, U.S.A.}

\begin{document}

\section{Introduction}

Although several astrophysical and cosmological observations strongly support the existence of Dark Matter (DM), its nature is yet to be understood~\cite{Bertone:2004pz,Bertone:2018xtm}. The standard $\Lambda$CDM cosmological model successfully explains the large-scale structure of the Universe and its formation through the collapse of primordial density fluctuations. The dominant contribution to the matter content of the Universe is due to the Dark Matter that is assumed to be cold (CDM) and almost collisionless. This means that at early times DM particles have non-relativistic velocities so allowing the structures to gradually form. Under these assumptions, ab-initio N-body cosmological simulations have pointed out that DM haloes are filled by DM substructures~\cite{Kuhlen:2012ft}. A key quantity is the subhalo mass function ${\rm d}N / {\rm d} \ln m_{\rm sub}$, that is the abundance of DM subhaloes per unit logarithmic subhalo mass $m_{\rm sub}$ within a host DM halo $M$ at a given redshift.

\begin{figure}[tbh!]
    \centering
    \includegraphics[width=0.5\textwidth]{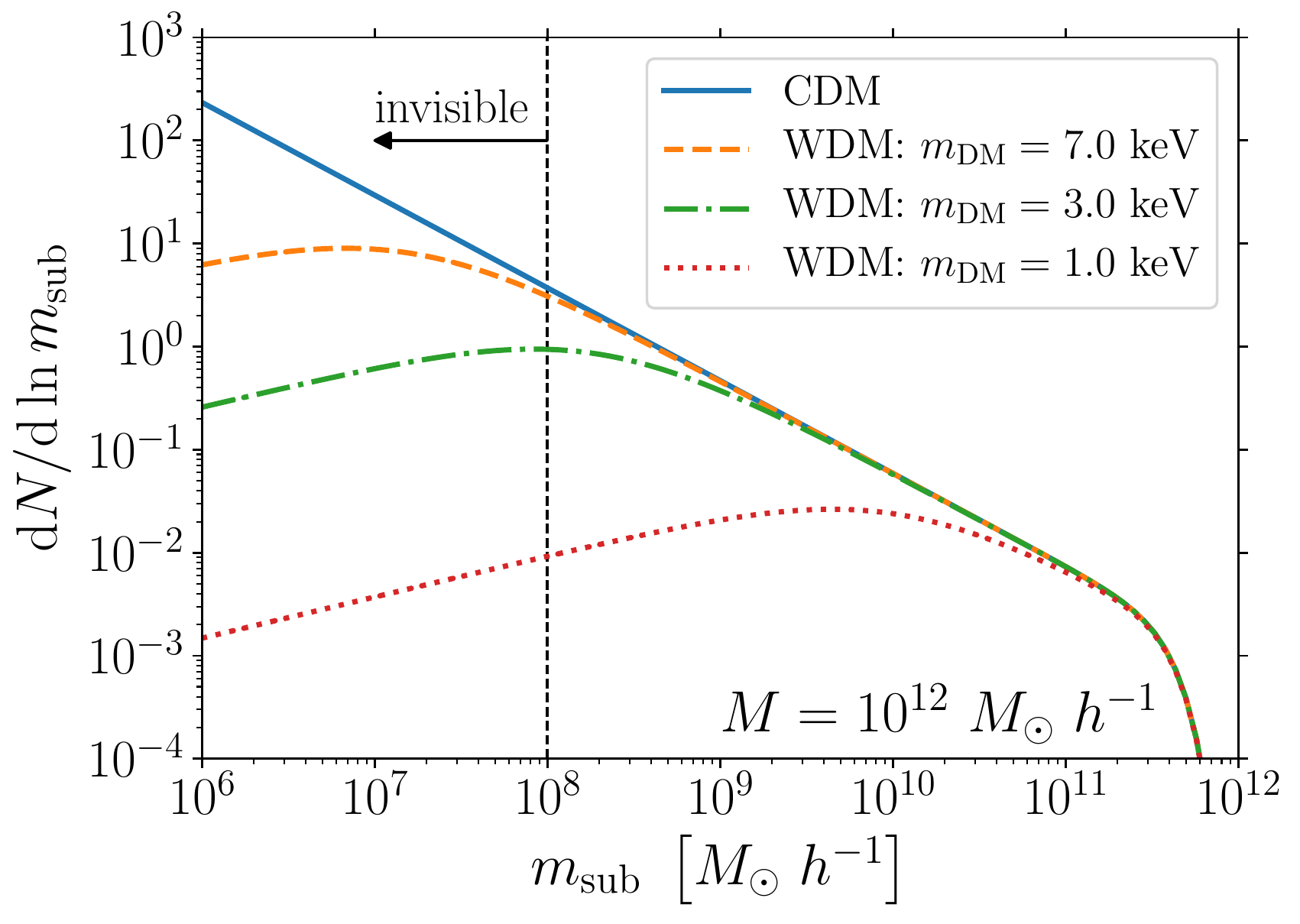}
    \caption{\textbf{Subhalo mass function from N-body simulations for CDM and WDM paradigms with different DM masses.} Subhalo mass function as a function of the subhalo mass within a host DM halo of $M = 10^{12}~M_\odot~h^{-1}$. The lines represent the expectation for CDM particles (solid blue lines) and WDM particles with different masses. Subhaloes with mass smaller than about $10^8~M_\odot~h^{-1}$ (vertical line) are expected to be dark due to the suppression of star formation.}
    \label{fig:1}
\end{figure}

A still viable alternative to the CDM paradigm is given by Warm Dark Matter (WDM). In this case, at the time of structure formation, WDM particles have instead non-negligible velocities that can wash out the primordial matter density fluctuations at about 1~kpc comoving scale. This results in a suppression of the formation of small-mass subhaloes. The free-streaming effects on the subhalo mass function are in general parameterized by the functional form~\cite{Lovell:2013ola}
\begin{equation}
    \frac{{\rm d} N_{\rm WDM}}{{\rm d} \ln m_{\rm sub}} = \frac{{\rm d} N_{\rm CDM}}{{\rm d} \ln m_{\rm sub}} \left(1 + \frac{m_{\rm hm}}{m_{\rm sub}}\right)^{-1.3} \,,
\end{equation}
where for thermally produced dark matter particles, the half-mode mass $m_{\rm hm}$ is equal to~\cite{Schneider:2011yu,Viel:2013apy}
\begin{equation}
    m_{\rm hm} = 10^{10}\left(\frac{m_{\rm DM}}{1~{\rm keV}}\right)^{-3.33} ~M_\odot~h^{-1}\,,
\end{equation}
with $m_{\rm DM}$ being the DM mass. By using the above expressions and the fitting formula reported in Ref.~\cite{Giocoli:2009ie}, in Fig.~\ref{fig:1} we show the subhalo mass function with a DM host halo of $M = 10^{12}~M_\odot~h^{-1}$, for CDM and WDM paradigms with different DM masses. The smaller the mass of thermal DM particles, the fewer the DM substructures with small mass. Hence, a naive way to test the nature of dark matter is to count DM subhaloes by looking for faint and ultra-faint dwarf galaxies inside the Milky Way or other galaxies. This procedure has indeed set a lower bound on the mass of thermal DM particles: $m_{\rm DM} > 1.5$~keV~\cite{Lovell:2013ola} and $m_{\rm DM} > 2.3$~keV~\cite{Polisensky:2010rw} depending on different assumptions. However, star formation in DM subhaloes with $m_{\rm sub} \lesssim 10^{8}~M_\odot~h^{-1}$ is expected to be suppressed by reionization, interaction with their host and self-quenching~\cite{Fitts:2016usl,Read:2017lvq}. An alternative way to detect small-mass subhaloes and, consequently, to discriminate between CDM and WDM is by looking for their gravitational effects in strong gravitational lensing systems~\cite{Vegetti:2012mc,Vegetti:2014lqa,Despali:2017ksx,Vegetti:2018dly,Ritondale:2018cvp}.

Strong gravitational lensing is a gravitational effect that allows us to observe multiple images of a background galaxy (the source) through the deflection of the light due to the distribution of matter along the line-of-sight (see Ref.~\cite{Treu:2010uj} for a review). Typically, the main contribution to the lensing comes from a foreground galaxy (the lens) that is almost aligned to the background one. In this case, we can observe the so-called Einstein ring. For these systems, the presence of DM subhaloes bounded to the lens galaxy and of small DM haloes along the line-of-sight (field haloes) affects the lensed images below the percentage level only. Hence, to detect these very small distortions, an accurate and fast analysis pipeline for strong gravitational lensing data is required.

In strong gravitational lensing analyses, the zeroth order step is to reconstruct the observed lensed image by modeling at the same time the surface brightness of the source galaxy and the matter distribution of the main lens galaxy. Regarding the source, two methods have been exploited so far: the use of parametric models like the Sersic profile or the pixelated reconstruction of the source~\cite{Koopmans:2005nr,Vegetti:2008eg}. While the former is not able to cover all the complex morphologies of galaxies, the latter has no {\it a priori} information of real galaxies and, therefore, requires a regularization according to some criteria (like smoothness among adjacent pixels). On the other hand, Machine Learning techniques can be employed to generate more realistic templates for the background source based on our knowledge and observation of real galaxies.

Deep learning methods have already been used for galaxy classification and for searches of strong lensing systems in huge observational catalogs of galaxies~\cite{2017MNRAS.472.1129P,2019MNRAS.482..807P}. Moreover, Convolutional Neural Networks (CNNs) have been proposed to quickly recover the main parameters describing the matter distribution of the lens (the size of the Einstein radius, the eccentricity, and so on)~\cite{Hezaveh:2017sht,PerreaultLevasseur:2017ltk}. Recently, recurrent inference machines have been shown to successfully reconstruct the background source, once they are coupled to an additional CNN that provides the lens model~\cite{Morningstar:2019szx}. Both pipelines are based on supervised learning, namely the CNNs are trained by feeding them with mock lensed images for which the lens and source true models are known. This does not seem to be a viable approach if one wants to include subhaloes and field haloes since this would highly enlarge the parameter space of the system and, consequently, the required training data set.

Differently from the aforementioned methods, we develop a new analysis pipeline that combines deep CNNs with physical models in order to increase the interpretability of the results~\cite{Chianese:2019ifk}. In particular, the surface brightness of the source galaxy is modeled by a Variational AutoEncoder (VAE), that is an unsupervised generative model consisting of two Neural Networks, the encoder and the decoder~\cite{2013arXiv1312.6114K,2014arXiv1401.4082J,DCGAN}. In particular, the encoder maps the input images to a latent space with a smaller dimensionality, while the decoder produces output images given the latent space variables. Once the VAE is trained with a data set of observed non-lensed galaxies, the decoder can be used to generate reasonable galaxy images. The lens model is instead described by physical models that depend on parameters with explicit physical meaning. The whole analysis pipeline is implemented in a differential probabilistic programming framework using Pytorch~\cite{paszke2017automatic} and Pyro~\cite{bingham2018pyro} packages. This allows us to automatically and efficiently compute the derivatives of the model with respect to all the source and lens parameters. These derivatives are used to find the best-fit parameters (optimization) and to estimate the parameter posteriors with a Bayesian approach by means of traditional sampling techniques such as Hamiltonian Monte Carlo (HMC)~\cite{Duane:1987de,2017arXiv170102434B}.

\section{Lensing analysis pipeline}
\begin{figure}[t!]
    \centering
    \includegraphics[width=0.75\textwidth]{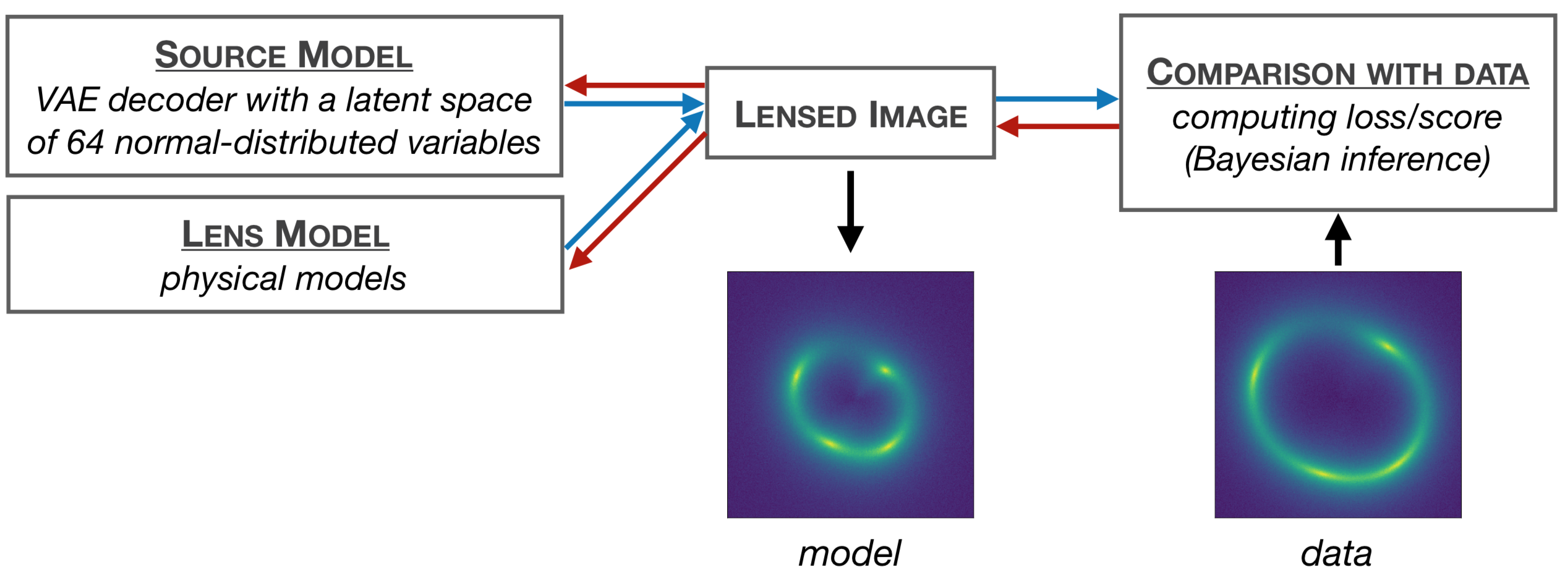}
    \caption{\textbf{Our analysis pipeline implemented in a differential probabilistic programming framework.} The forward model (blue lines) computes the lensed image combining the source model (the VAE decoder) and the lens model (physical models for the lens mass distribution). Then, the image is compared with the data to infer the model parameters in a Bayesian approach. The backward model (red line) computes the derivatives of the forward model with respect to all the parameters.}
    \label{fig:2}
\end{figure}

The lensing pipeline is depicted in Fig.~\ref{fig:2} (see Ref.~\cite{Chianese:2019ifk} for more details). The lensed image is obtained by solving the lens equation on a squared pixel grid under the thin lens approximation. The lens equation defines a relation between the two-dimensional angular coordinates in the lens plane $\bm{\theta}_{\rm lens}$ and the ones in the source plane $\bm{\theta}_{\rm src}$, that is
\begin{equation}
    \bm{\theta}_{\rm src} = \bm{\theta}_{\rm lens} - \bm{\alpha}\left[\Sigma\right]\,,
    \label{eq:lens_eq}
\end{equation}
where $\bm{\alpha}$ is the displacement field that depends on the projected surface mass density of the lens, here denoted as $\Sigma$. Since the surface brightness $I$ (photon flux density per unit angular area) is preserved by the lens equation, the lensed image is simply given by
\begin{equation}
    I(\bm{\theta}_{\rm src}) = I(\bm{\theta}_{\rm lens} - \bm{\alpha}\left[\Sigma\right])\,,
\end{equation}
The two main ingredients are therefore the surface brightness $I$ of the source galaxy and the mass distributions $\Sigma$ along the line-of-sight, which are computed in the source model and the lens model, respectively (see Fig.~\ref{fig:2}).

The lens model is completely described by semi-analytical expressions that are based on physical models. The displacement field is computed on a squared pixel grid in the lens plane as the sum of two contributions: a main halo parameterized as a Singular Power-Law Ellipsoid (SPLE) profile~\cite{1993ApJ...417..450K} and an external shear. The SPLE has six parameters: the Einstein radius $r_{\rm Ein}$, the slope of the profile $\gamma$, the $x$ and $y$ coordinates of the center, the major-to-minor axis ratio and the position angle. The external shear contribution is defined by two additional parameters. Dark matter subhaloes and field haloes will be included and discussed in a future work.

The source model is given by the decoder of a trained Variational AutoEncoder with a latent space of normal-distributed 64 variables $z$ (the architecture will be described in detail in Ref.~\cite{Chianese:2019ifk}). The VAE is trained by means of the GREAT3 catalog that contains more than $1.5 \times 10^4$ observed galaxies~\cite{Mandelbaum:2013esa}. The training procedure is required to estimate the parameters (the weights and the biases) of the convolutional and deconvolutional neural networks (the encoder and the decoder, respectively) and the prior distributions of the latent space. Such distributions are used to sample the latent variables $z$ in order to reproduce realistic galaxy images through the decoder. The output galaxy is defined on a squared pixel grid equal to the one defined in the lens plane. Hence, to compute the surface brightness in any position of the source plane we consider a bilinear interpolation among adjacent pixels.

The final lensed image is obtained by applying a symmetric two-dimensional Gaussian point spread function and adding uncorrelated Gaussian noise to each pixel. Then, it is compared with the data quantifying with a score of how much the two images differ each other. The loss function depends on the inference procedure: for example, it can be the pixelwise mean squared error or the so-called evidence lower bound. So far, we have described the forward model (blue arrows in Fig.\ref{fig:2}), that is the computation of the loss function given the parameters of the source and lens models. Then, the differential probabilistic programming framework allows us to compute derivatives of the loss function with respect to all the parameters. This is defined as backward model (red arrows in Fig.\ref{fig:2}).

The analysis pipeline is defined by two steps: the optimization and the inference through a Bayesian approach. During optimization, starting from a random initial configuration, the best-fit parameters are found by minimizing the loss function. The parameters are updated by means of gradient descent. For this step, we perform a maximum a posteriori probability estimation (MAP inference) by considering delta-Dirac distributions for each parameter. As soon as the optimization converges (the loss becomes almost constant), we compute the posterior distributions of the parameters starting from the best-fit points. We use the HMC sampling technique, so taking advantage of the gradient descent algorithm once again.

\section{Source and Lensed Images Reconstruction}
\begin{figure}[t!]
    \centering
    \includegraphics[width=1.0\textwidth]{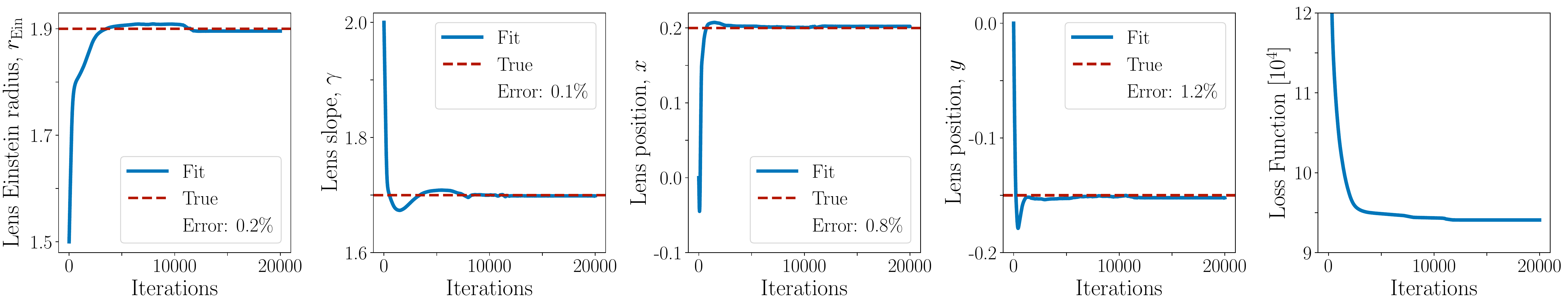}
    \caption{\textbf{Optimization procedure to find the best-fit points.} The main lens parameters and the loss function as a function of the number of iterations. The dashed red lines are the corresponding true parameters.}
    \label{fig:3}
\end{figure}

Here, to test our pipeline, we analyze a mock lensed image with 256$\times$256 pixels. The mock source galaxy corresponds to a denoised galaxy of the GREAT3 data set that was not included in the VAE training. In Fig.~\ref{fig:3}, we report some parameters of the main lens halo and the loss function (last panel on the right) as a function of the number of iterations during the optimization procedure. As one can see, the convergence is achieved after much less than $2\times 10^4$ iterations. Remarkably, the best-fit points are within $1\%$ interval with respect to the true values, which are represented in the plots by horizontal dashed red lines.
\begin{figure}[t!]
    \centering
    \includegraphics[width=0.7\textwidth]{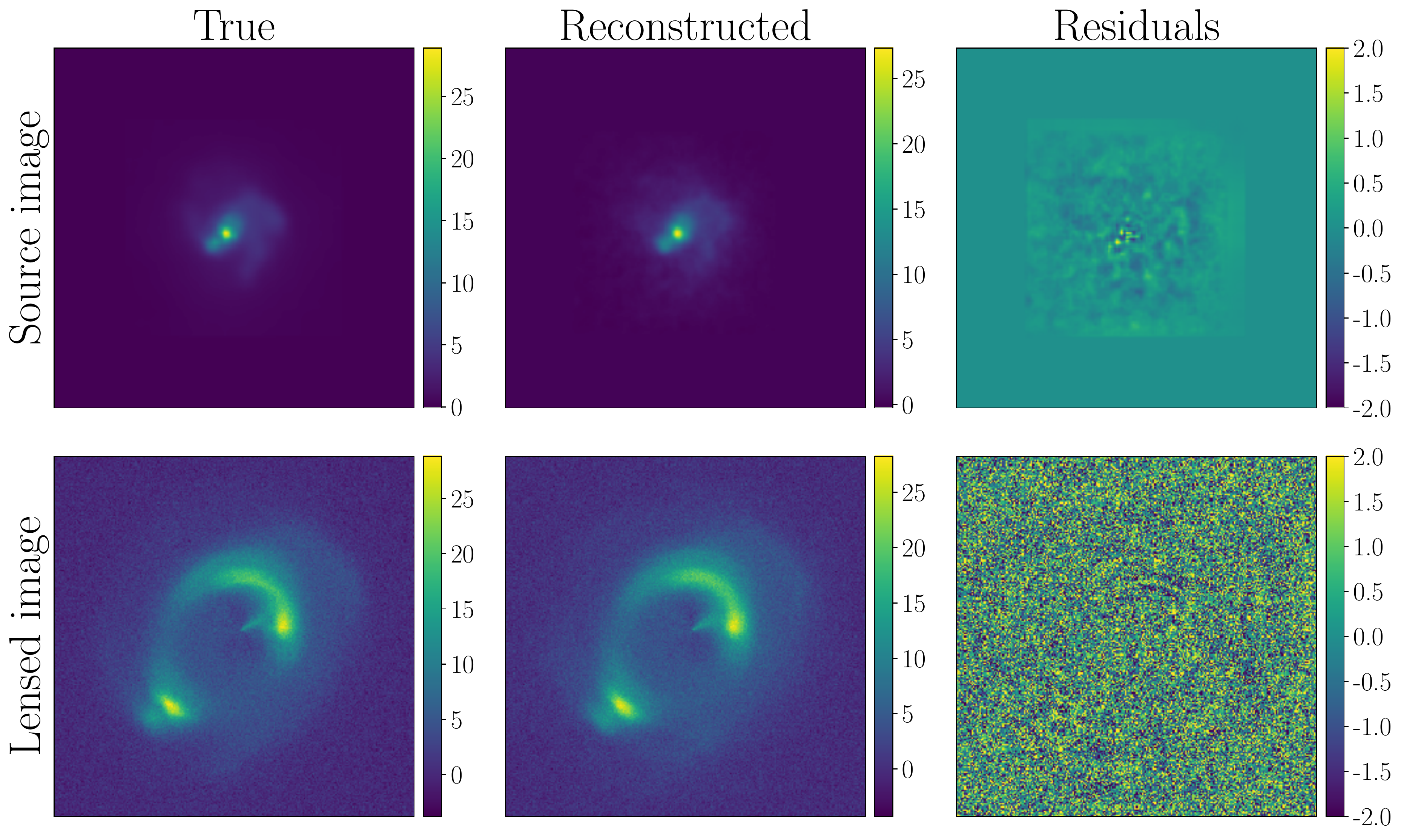}
    \caption{\textbf{Reconstruction of source and lensed images after the optimization.} The true source and lensed images (first column) are compared with the reconstructed ones (second column) by computing the corresponding residuals (third column).}
    \label{fig:4}
\end{figure}

The reconstructed source and lensed images are shown in Fig.~\ref{fig:4} (second column) and compared with the true ones (first column). The residuals are displayed in the third column. The reconstructed lensed image matches the original one at the noise level as highlighted by the residuals. On the other hand, the residuals in the source plane are higher, though the two source galaxies are qualitatively in agreement. For instance, the spiral arms are captured in the reconstruction.
\begin{figure}[t!]
    \centering
    \includegraphics[width=0.8\textwidth]{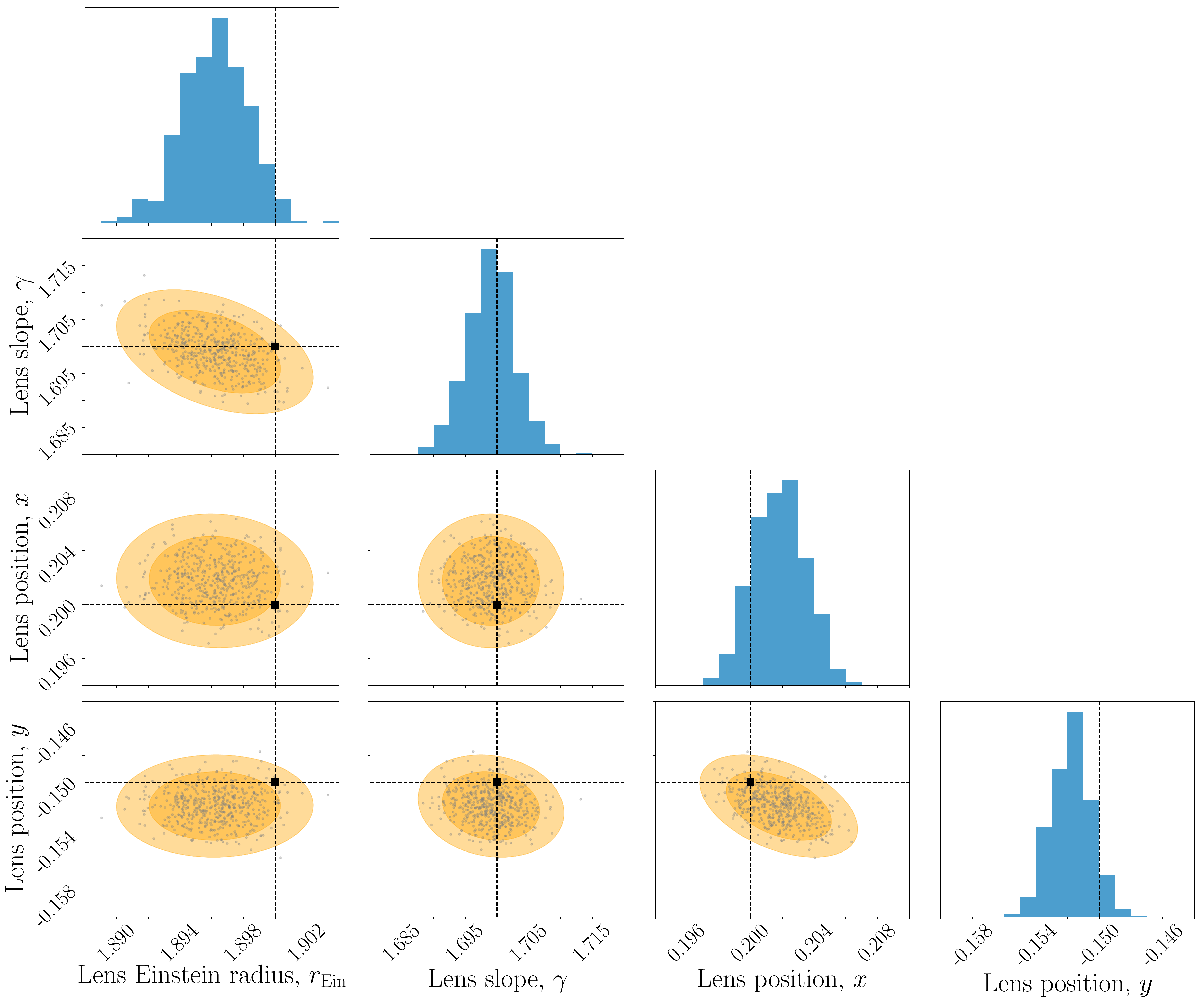}
    \caption{\textbf{Bayesian inference with Hamiltonian Monte Carlo.} The marginalized posterior distributions of the main lens parameters obtained by sampling 500 points with Hamiltonian Monte Carlo. The ellipses represents the contours at $68\%$ and $95\%$ levels. The true parameters are highlighted by the black squares.}
    \label{fig:5}
\end{figure}

Finally, in Fig.~\ref{fig:5} we report the posteriors of the main lens parameters obtained with the HMC starting from the best-fit points. These distributions are marginalized with respect the other lens and source parameters. The ellipses show the contours containing $68\%$ and $95\%$ of the 500 points (grey dots) provided by the HMC. The black squares represent the true parameters.

\section{Conclusions}

In this paper, we have briefly described the main ingredients of a new analysis pipeline for strong gravitational lensing data. The novelties of our approach are:
\begin{itemize}
    \item the use of a Variational AutoEncoder to generate the background source galaxy;
    \item the combination of deep learning techniques with physical models to increase the interpretability of the results and to make possible further generalizations of the lens model (for example, the inclusion of dark matter substructures);
    \item the differential probabilistic programming framework to automatically perform the Bayesian inference with gradient descent techniques.
\end{itemize}
The preliminary analyses of mock data are promising. In particular, we are able to infer the true parameters of the lens model at high accuracy (typically below the $1\%$ level). Moreover, the true parameters reside within the $95\%$ contours of the posterior distributions obtained by means of Hamiltonian Monte Carlo. However, further improvements are required on the VAE source model in order to reconstruct more complicated galaxies and reduce the residuals in the source plane. This will be discussed in detail in Ref.~\cite{Chianese:2019ifk}, where a more systematic study of the pipeline performance is also presented.

\end{document}